\documentstyle[aas2pp4]{article}

\newcommand{\beq}{\begin{equation}}
\newcommand{\eeq}{\end{equation}}

\def\beqa{\begin{eqnarray}}
\def\eeqa{\end{eqnarray}}
\newcommand{\lsim}{\lesssim}
\newcommand{\gsim}{\gtrsim}

\begin{document}
\noindent Accepted ApJ, 569, (April 20), 2002
\title{Squelched Galaxies and Dark Halos}

\author{R. Brent Tully$^{1}$, Rachel S. Somerville$^{2}$, Neil Trentham$^{2}$,
and Marc A. W. Verheijen$^{3,4}$}
\affil{{}$^1$Institute for Astronomy, University of Hawaii,
Honolulu, HI 96822\\
{}$^2$Institute of Astronomy, Cambridge University,
Cambridge, CB3 0HA UK\\
{}$^{3}$National Radio Astronomy Observatory, Socorro, NM 87801\\
{}$^{4}$Department of Astronomy,University of Wisconsin, Madison, WI 53706\\
}

\begin{abstract}
There is accumulating evidence that the faint end of the galaxy luminosity 
function might be very different in different locations.  The luminosity 
function might be rising in rich clusters and flat 
in regions of low density.  If galaxies form according to the model of
hierarchical clustering then there should be many small halos compared to
the number of big halos.  If this theory is valid then there must be a
mechanism that eliminates at least the visible component of galaxies in
low density regions.  A plausible mechanism is photoionization of the
intergalactic medium at a time before the epoch that most dwarf galaxies form
in low density regions but after the epoch of formation for similar systems 
that ultimately
end up in rich clusters.  The dynamical timescales are found to 
accommodate this hypothesis in a flat universe with $\Omega_m \lsim 0.4$.

If small halos exist but simply cannot be located because they have never
become the sites of significant star formation, they still might have
dynamical manifestations.  These manifestations are hard to identify in 
normal groups of galaxies because small halos do not make a significant
contribution to the global mass budget.  However, it could be entertained
that there are clusters of halos where there are {\it only} small systems, 
clusters that are at the low mass end of the hierarchical tree.  There may be
places where only a few small galaxies managed to form, enough for us to
identify and use as test probes of the potential.  It turns out that such
environments might be common.  Four probable groups of dwarfs are identified
within 5~Mpc and the assumption they are gravitationally bound suggests
$M/L_B \sim 300 - 1200~M_{\odot}/L_{\odot}$, 6 $\pm$~factor~2 times higher 
than typical
values for groups with luminous galaxies.

\end{abstract} \vspace{1 mm} 

\keywords{cosmology: dark matter -- galaxies: formation
--- galaxies: luminosity function, mass function}
\vspace{3 mm}\

\twocolumn

\section{Expectations}

According to the popular cold dark matter (CDM) hierarchical clustering model 
of galaxy formation
there should be numerous low mass dark halos still around today.  The 
approximation by \cite{PS74} (1974) that initial density fluctuations
would grow according to linear theory to a critical density and then
collapse and virialize leads, with a CDM-like power spectrum, to a prediction 
of sharply increasing numbers of
halos at smaller mass intervals.  Cosmological simulations are now
being realized with sufficient mass resolution to distinguish dwarf galaxies 
and this modeling basically confirms expectations of the existence of 
low mass halos (\cite{KKVP99} 1999; \cite{M+99} 1999).  
Ninety percent of low mass halos accreted into a cluster may be disrupted by
tidal stripping or absorbed by dynamical friction, but the halo mass function
is still anticipated to rise steeply toward lower masses (\cite{BKW00} 2000).

Indeed, dwarf galaxies are found in abundance in some environments.  
In the past, most observational effort has gone into studies in rich clusters 
because the
statistical contrast is highest against the background (\cite{SDP97} 1997;
\cite{T98} 1998; \cite{PPSJ98} 1998; also the small but dense Fornax Cluster:
\cite{KDSBH00} 2000).  The general conclusion from these
studies has been that, yes, there are substantial numbers of dwarfs of the
spheroidal type.  The high dwarf fraction reported in some instances may be in
 agreement with expectations
of CDM hierarchical clustering theory.

However, there has been a suspicion that there might not be the expected 
abundance of dwarfs in environments less extreme in density than the rich 
clusters.  \cite{KKVP99} (1999) and \cite{M+99} (1999) have pointed out the 
apparent absence of
large numbers of dwarfs in the Local Group.  It is to be appreciated that
the task of identifying extreme dwarfs is not trivial.  They are tiny and
faint.  At substantial distances their surface brightnesses are faint against
the sky foreground and close up they resolve into swarms of very faint stars.
So dwarfs were not being found in the expected numbers but is this because
of observational limitations?

Already at relatively high intrinsic luminosities there is good evidence of
variations of the galaxy luminosity function with environment.  The luminosity
function is steeper (larger dwarf/giant fraction) in denser groups
characterized by thermal X-ray emission or high velocity dispersions
(\cite{ZM00} 2000; \cite{C00} 2000).  The trends are subtle in these
studies because the faint end cutoffs barely include what would normally be 
considered dwarf galaxies.  For example, \cite{ZM00} go comparatively faint,
to $M_R=-16.6+5{\rm log}h_{75}$, where $h_{75}={\rm H}_{\circ}/75$.

\section{Four Environments}

Motivated by the speculation that the occurrence of dwarfs might be correlated
with local density, we made extensive observations in the nearest environment
where the density is low (dynamical time is long) yet where there are enough
galaxies for a meaningful statistical discussion.  We studied the Ursa Major
Cluster, a structure fortuitously at about the same distance as the Virgo 
Cluster and which subtends a comparable amount of sky.  The total light
in bright galaxies in Ursa Major is about 1/4 that in Virgo but dynamical 
evidence suggests that the mass in Ursa Major is down by a factor 20 from 
that associated with Virgo (\cite{TS98} 1998).  Roughly 16 sq. deg. of the Ursa
Major Cluster 
were surveyed with deep CCD imaging with wide field cameras on the 
Canada-France-Hawaii Telescope and in the 21cm Hydrogen line with the 
Very Large Array. 
Results of the two aspects of the survey are being 
reported respectively by \cite{TTV01} (2001) and \cite{VTTZ00} (2000 and in
preparation).  The optical survey provides information on dwarfs to a 
completeness limit of $M_B=-10$.  The HI survey confirms that there is no
hidden component that is gas rich.

In a separate study, \cite{TH02} (2002) have searched for dwarfs in a 25 sq.
deg. swath of the Virgo Cluster using images acquired as part of the  2.5m
Isaac Newton Telescope Wide Field Survey.  Roughly 20\% of the cluster is
covered and provides information on the luminosity function down to a
completeness limit $M_B \sim -11$.  The results are consistent with what was
found by \cite{SBT85} (1985) but quite divergent from the situation suggested
by \cite{PPSJ98} (1998).  These latter authors found a pronounced steepening
of the luminosity function faintward of $M_R=-15.5$ ($M_B \sim -14$) which
\cite{TH02} speculate is attributed to background contamination.
In any event, as we will see further along, whether the moderately rising
luminosity function of \cite{SBT85} and \cite{TH02} is a correct description,
or the more extreme situation described by  \cite{PPSJ98} is correct, the
faint end of the galaxy luminosity function is steeper in the Virgo Cluster
than in the Ursa Major Cluster.

Two more extreme environments have reasonably well defined luminosity 
functions: the high density Coma Cluster and the low density Local Group.  The 
Coma Cluster
has been studied by \cite{T98a} (1998).  The data assembled for the Local Group
is discussed by \cite{TH02}.  The $B$-band luminosity function data for all
four environments discussed above are shown in Figure~1 (extracted 
from \cite{TH02}).

\begin{figure}[htb]
\plotone{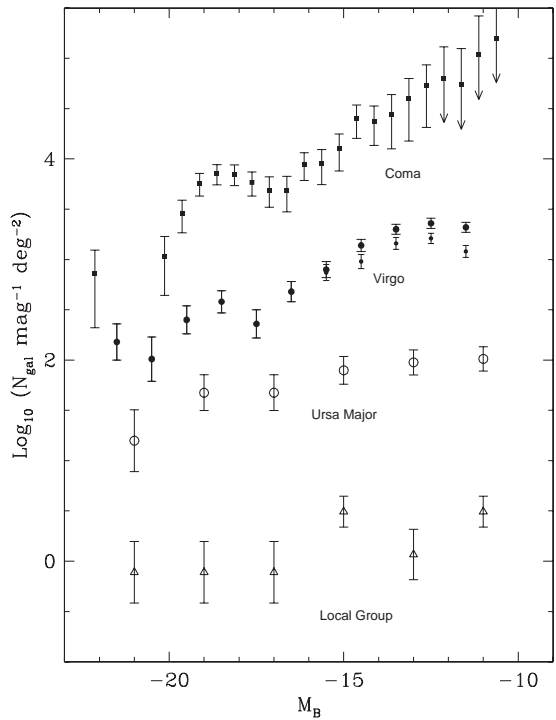}                     
\caption{
$B$-band luminosity functions.  From the top, Coma Cluster (filled squares),
Virgo Cluster (filled circles), Ursa Major Cluster (open circles), and the
Local Group (open triangles).  Vertical scales are shifted for clarity.
Downward arrows in the case of the Coma Cluster reflect background 
contamination uncertainties. 
\label{1} }
\end{figure}

The main qualitative point to be drawn from this figure is that the slopes of
the four luminosity functions are {\it not} the same.  There is a steepening
correlated with the richness of the cluster, progressing from the Local Group,
through Ursa Major Cluster, the Virgo Cluster, to the Coma Cluster.
The observational situation is still not fully clear
but the case is becoming strong that there are environmental
differences in the {\it opposite} sense of the expected
dark matter halo mass trends that are discussed in the next section.

\section{Squelched Galaxies}

According to hierarchical clustering theory, as time goes on
small halos merge or are disrupted.  Still, the theory 
anticipates that there should be numerous
dwarf galaxies relative to giant galaxies, perhaps in accordance with the
observed situation in rich clusters.  The ratio of dwarfs to giants is 
expected to depend on the pace of the merging process which is governed
by the local density.
At first thought, it would seem that
the rich clusters are more hostile, the low density regions more benign for
the survival of small galaxies.  In very low density groups dynamical 
collapse times can be of order the age of the universe.  The merging
process is not so far advanced and
dynamical friction and tidal stripping have reduced consequences.
In N-body simulations, \cite{S00} (2000) have found a greater survival of 
small halos compared to large ones in more isolated environments.
So the general expectation would be that there are more dwarfs per giant
in low density, less evolved regions, precisely the opposite of what is
indicated in Fig.~1.
Conceivably, galaxy harassment of some sort could transform
giant galaxies into dwarfs in clusters (\cite{MLK98} 1998).  However, it is
in clusters where there is better agreement between observations and 
theoretical expectations.  We are confronted with the problem of the 
{\it absence} of dwarfs in low density regions.  If dwarfs can be formed
from harassment in clusters it
would only extend the disparity with the predictions of CDM theory to clusters
(the expected large numbers of primordial dwarfs would not be found anywhere).
At face value, we need to call upon a mechanism that {\it allows} small 
galaxies 
to form in rich clusters but {\it thwarts} small galaxy formation in places 
of low density.

A plausible mechanism is 
photoionization of the intergalactic medium before the epoch of galaxy
formation.  \cite{E92} (1992) discussed the inhibiting
effect on the formation of dwarfs due to the suppression of cooling of a 
primordial plasma of hydrogen and helium.  \cite{TW96} (1996) took the 
discussion further with recourse to high resolution hydrodynamic simulations.
These authors argue that gas heating before collapse is more important than
inhibition of line cooling.  The suppression of galaxy formation occurs
below a virial velocity threshold.  The UV background heats the precollapse gas 
to roughly 25,000~K.  This temperature is much less than that associated with
the virial energy of a large galaxy, hence has negligible effect on the 
collapse of baryons into a massive potential well.  However, for a sufficiently
small galaxy this heating is comparable with, or can dominate, the 
gravitational energy.  \cite{TW96} and also \cite{G00} (2000) find there is 
essentially total suppression
of baryon collapse for systems with circular velocities 
$V_{circ}\lsim30$~km~s$^{-1}$ and, by contrast, little effect on galaxy 
formation for systems with $V_{circ}\gsim75$~km~s$^{-1}$.  It follows that
luminosity functions would be little affected above 
$M_B\sim-18+5{\rm log}h_{75}$ but strongly attenuated below 
$M_B\sim-15+5{\rm log}h_{75}$. 

The suppression of baryon collapse would only apply to
galaxy formation that occurs after 
reheating of the
intergalactic medium.  The collapse timescale (\cite{GG72} 1972) is
\beq
t_{col}=1.4\times10^{10}(R_{vir}^3/M_{14})^{1/2} h_{75}^{-1}~{\rm yr}
\label{tcol}
\eeq
where $R_{vir}$ is the virial radius in Mpc and $M_{14}$ is the virial mass in 
units of $10^{14}~M_{\odot}$.  
Values for $R_{vir}$ and $M_{14}$ can be extracted from
\cite{T87} (1987) for the Virgo and Ursa Major clusters ($R_{vir}$: 0.79 and 
0.98~Mpc
respectively; $M_{14}$: 8.9 and 0.5 respectively).  Hence, rough dynamical
collapse times for these clusters are $t_{col}^{virgo}\sim3.3$~Gyr and
$t_{col}^{uma}\sim19$~Gyr.   The dense, elliptical dominated Virgo Cluster 
formed a {\it core} 
long ago and the loose, spiral dominated Ursa Major Cluster is still in the
process of collapsing.  Of course, galaxies continue to fall in and enlarge
the Virgo Cluster to this day and, on the other hand, 
substructure in Ursa Major would have shorter dynamical collapse times than
the entire entity.

Smaller mass scales collapse before larger mass scales.  Dwarfs must form
before their host cluster forms.  The timing of halo
collapse and mergers as a function of environment will be considered in the 
next section.  To 
conclude this section, we review the evidence on the timing of
reionization of the
intergalactic medium by the UV radiation of AGNs or hot stars.  

Observations
constrain the epoch of reionization to $z\gsim6$ (\cite{F+00} 2000; 
\cite{Be+01} 2001), which can be
understood on theoretical grounds (\cite{GO97} 1997).
In Figure~2 we see the relationship between
redshift and the age of the universe for a wide range of topologically flat
cosmological models.  If baryon collapse into small galaxies can only occur
before reionization then Fig.~2 tells us crudely that if the epoch of 
reionization 
is as late as $z_{ion}\sim6$ then dwarfs with $t_{col}\sim1$~Gyr could 
form in a universe with 
matter density $\Omega_m\sim0.2$ and vacuum energy density 
$\Omega_{\Lambda}\sim0.8$.

\begin{figure}[ht!]
\plotone{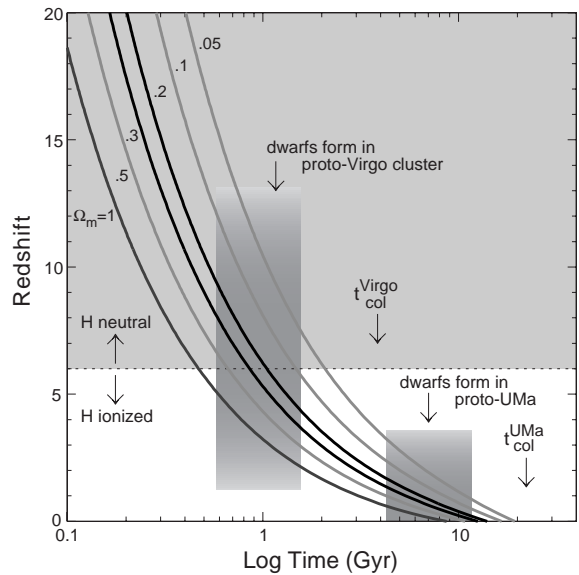}
\caption{Redshift vs age of the universe for a range of flat world models,
from $\Omega_m=1$, $\Omega_{\Lambda}=0$ on the bottom to 
$\Omega_m=0.05$, $\Omega_{\Lambda}=0.95$ on top.  The arrows
indicate the rough epochs of galaxy formation in the Virgo and Ursa Major
clusters and the collapse timescales of the clusters.  Intergalactic 
reionization must have occurred at $z_{ion}\gsim6$; above
the horizontal dotted line.}
\end{figure}

\section{Simulations}

We use semi-analytic models of galaxy formation based on the code developed
by \cite{S97} (1997) and described in detail by \cite{SP99} (1999) and 
\cite{SPF01} (2001).
The formation history of collapsed dark matter halos and their sub-structure
is described via Monte Carlo ``merging trees'' based on the extended
Press-Schechter formalism (\cite{SK99} 1999).  Radiative cooling by atomic 
Hydrogen, formation of stars, and feedback due to supernovae winds is 
modeled with simple empirical recipes, described in the references above.
In this paper, 
we also include a recipe to suppress gas accretion because of heating
by an external photo-ionizing background, a feature not previously
included in a full semi-analytic model.  We describe this new ingredient 
briefly below.  The concepts are discussed further by \cite{S01} (2001).

A recipe for suppression of gas collapse is adopted from \cite{G00} (2000) and
produces results consistent with \cite{TW96} (1996). 
Reionization is assumed to take place instantaneously at a redshift $z_{ion}$. 
In halos of virial mass $M_{vir}$ that collapse before reionization, the mass
of participating gas $M_g$ available to ultimately make stars is:
\beq
M_g = f_b M_{vir}
\label{mhotb}
\eeq
where $f_b=\Omega_b/\Omega_m$ is the universal baryon fraction.
In halos collapsing after reionization, there is suppression of the 
participation of gas in the collapse:
\beq
M_g = {f_b M_{vir}\over[1+0.26 M_{50}/M_{vir}]^3}
\label{mhota}
\eeq
where halos with $M_{50}$ retain 50\% of their baryon mass.  Acceptable results 
are 
found if $M_{50}(z_c)$ is the mass associated with a halo with virial velocity
50~km~s$^{-1}$ at a collapse epoch $z_c$.  Since halos collapsing later have
lower density, $M_{50}(z_c)$ increases as $z_c$ decreases.
It follows from this recipe that,
after reionization, gas collapse is suppressed completely in halos
with $V_{cir} < 30$~km~s$^{-1}$ and is almost unaffected in halos with 
$V_{cir} > 75$~km~s$^{-1}$.  When this new ingredient is included, the
luminosity function of satellite galaxies in the Local Group predicted by our
model is in good agreement with observations (\cite{S01} 2001).

When do dwarf galaxies form in environments of different total mass?  This
question can be addressed by following back the merger trees in semi-analytic
simulations.  As a matter of definition, it is taken that a sub-halo within a
parent halo forms at the
redshift $z_f$ when the largest progenitor has a mass of half the final 
sub-halo mass.  This discussion
considers only final sub-halos with virial velocities in the range 
$17<V_{cir}<50$~km~s$^{-1}$, the range strongly susceptible to squelching of
star-formation by reionization.  

The merger trees can be traced back in virialized parent halos with a range
of masses, $M_H$.  In Figure~3, we see the distribution of formation redshifts
for squelchable dwarf halos embedded in parent halos with masses 
$10^{11}-10^{14}M_{\odot}$.  The solid histograms are based on the `progenitor
with half the
final mass' definition of sub-halo formation while the dotted histograms 
represent
the formation epoch of the `oldest progenitor' (the redshift at which the first
progenitor has gas at $10^4$~K that can cool).
The quantity $dP/dz_f$ is the fraction of dwarf 
halos with formation redshifts in the interval $z$, $z+dz_f$.

\begin{figure}[ht!]
\plotone{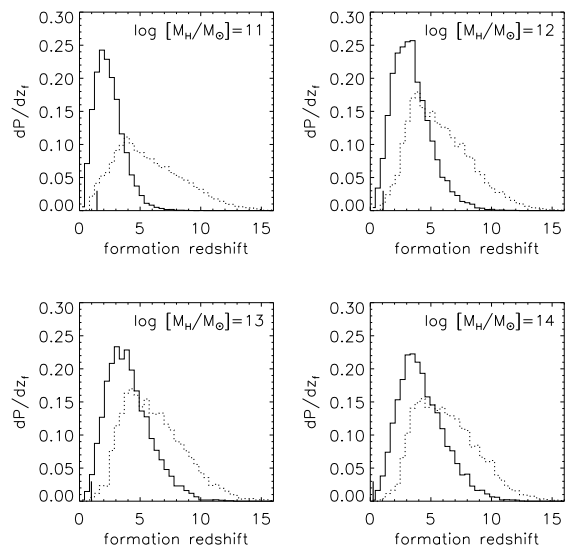}
\caption{Distribution of formation redshifts for dwarf halos 
($17<V_{cir}<50$~km~s$^{-1}$) within virialized parent halos ranging from 
$10^{11}M_{\odot}$ to $10^{14}M_{\odot}$.  Solid histograms: formation epoch 
defined by development of a progenitor with half the final dwarf sub-halo mass.
Dotted histograms: formation epoch of first progenitor to cool.  
The vertical tick marks in the lower left corners of each panel indicate the
average formation redshift of the parent halos (half the final mass in place).
More massive halos formed later but, within them, small halos tended to form 
earlier in environments that became more massive clusters.
}
\end{figure}

The definition of the formation epoch in terms of the development of a
progenitor with 50\% of the final mass is arbitrary.  
Figure~4 shows the cumulative distribution of the fraction of the final
sub-halo mass that is in a single progenitor at $z=8$; i.e. if we define
$f \equiv M(z=8)/M_0$ where $M_0$ is the final mass of the dwarf sub-halo, 
then the plot shows the fraction of objects
whose largest progenitors have fractional mass greater than the quantity
plotted on the x-axis. Thus, if we assume a simple picture in which a
sub-halo survives squelching if it has some fraction of its final mass in
place at reionization (as in the model of \cite{B+01} 2001), then we
can read the fraction of surviving galaxies off of the plot for any
assumed value of the critical fraction $f$ and for various parent halo
masses. We see again that a much larger fraction of dwarfs will survive
squelching in high-mass halos. The plot assumes $z_{ion} = 8$ but
the results are insensitive to the precise epoch of reionization.

\begin{figure}[ht!]
\plotone{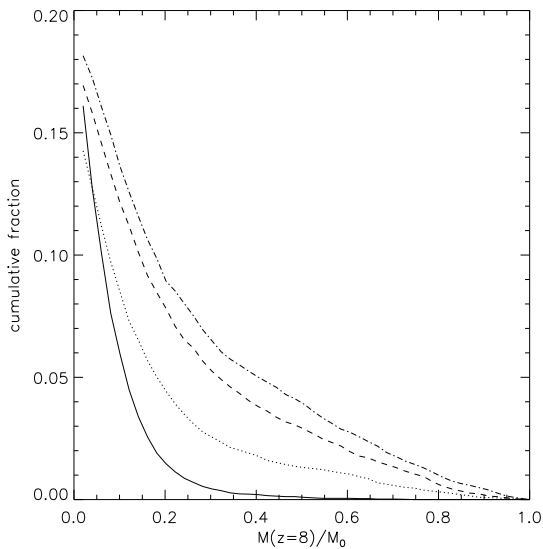}
\caption{Cumulative fraction of sub-halos of ultimate mass $M_0$ in place by 
$z=8$.  The four curves differentiate between parent halos with a range of mass:
$M_H=10^{11} M_{\odot}$ (solid curve), $M_H=10^{12} M_{\odot}$ (dotted curve),
$M_H=10^{13} M_{\odot}$ (dashed curve), and $M_H=10^{14} M_{\odot}$ (dash-dot 
curve).
}
\end{figure}

The quantity $dP/dz_f$ shown in Fig.~3 can be integrated to determine the 
fraction of dwarf halos that formed
before the epoch of reionization $z_{ion}$ in any specified parent halo
$P_{z_{ion}}(M_H,z_f>z_{ion})$.
This quantity is the fraction of dwarf halos amenable to the collection of cold
gas and hence the formation of a visible galaxy.  Values for $P_{z_{ion}}$ are
shown as a function of parent halo mass in
Figure~5.

\begin{figure}[ht!]
\plotone{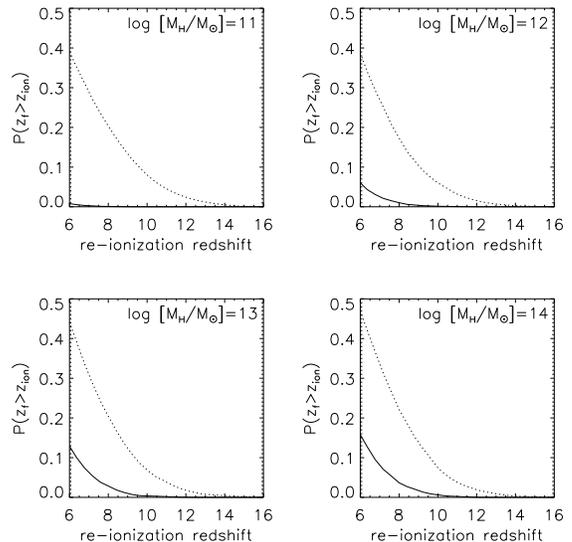}
\caption{Fraction of dwarf halos formed before reionization in parent halos
ranging in mass from $M_H=10^{11} M_{\odot}$ to $M_H=10^{14} M_{\odot}$.
Solid curves: galaxy formation epoch defined by acquisition of half the final
mass.  Dotted curves: epoch of oldest progenitor.
}
\end{figure}

Two clear conclusions can be drawn from this figure.  First, whatever the
epoch of reionization, the fraction of dwarf halos that can accumulate cold
gas before reionization
is greater in more massive parent halos.  That is, dwarf halos formed
earlier in environments that become massive clusters.  Second, whatever the
parent halo mass, the mechanism of star formation squelching by reionization
is more effective the larger the redshift of reionization.  Inverting this 
last  point, the trend of the dwarf fraction with mass is expected to be
stronger if reionization is later.

Qualitatively, it is plausible that the larger dwarf fraction in the Virgo
Cluster ($8 \times 10^{14} M_{\odot}$) comes about because many dwarf halos
were in place in the proto-Virgo region before reionization, while the
smaller dwarf fraction in the Ursa Major Cluster ($4 \times 10^{13} M_{\odot}$;
or smaller since U Ma is probably not virialized at this mass)
is a consequence of the fact that few dwarf halos were in place before
reionization.
Interestingly, this squelching mechanism only produces a pronounced
differential with environment in a universe with relatively low
matter density, say $\Omega_m<0.4$, $\Omega_{\Lambda}>0.6$.  In a universe
with $\Omega_m=1$, structure forms at low redshift: $t_{col}\sim1$~Gyr 
corresponds to $z\sim3$.

It would follow that if a range of cluster environments is explored then
there should be a roll-over: denser clusters with short dynamical times will
have a large dwarf/giant fraction and less dense clusters with long dynamical 
times will have a small dwarf/giant fraction.
The collapse time scale associated with a break point density
would reflect the time of reionization of the universe.

In addition to counts of numbers of dwarfs, it is known that the fractional 
representation of different types of dwarfs depends on environment 
(\cite{BST88} 1988).
There are more gas-depleted types in denser clusters, more irregulars with
ongoing star formation in low density regions.  If the halos of all these 
systems
were in place and gas had accumulated before the time of reionization
(our hypothesis) then
it would follow that there would be very old stars in these systems, whatever
their type.  In fact, old stars are found in all galaxies that have been 
observed appropriately.  Evidently the pace of the transition of cold gas
into stars is highly variable from dwarf to dwarf.  Even among spheroidals
that have no gas or recent star formation, there is evidence in well studied
nearby cases that star formation went on in spurts over extended intervals
(\cite{M98} 1998).

\section{A Search for Dark Halos}

If the preceding ideas have any merit then it follows that there would be
many low mass halos that are not identified because they contain few stars
or neutral gas.  In groups with luminous galaxies, the {\it fractional}
mass representation of unlit halos would be expected to be small and such
halos would have little dynamical consequence.
The overall ratio of mass to light would 
be reflective of the properties of the dominant galaxies and their halos.
In the very low mass regime, maybe there are collapsed halos where {\it no} 
stars were born, but we cannot find those places.  In between, following 
from the hypothesis that there are squelched dark halos, there could be
groups where {\it some} halos have birthed stars but where $M/L$ is large.
Maybe there might be groups with {\it only} low luminosity galaxies where
the contribution to the mass inventory from dark halos might be 
significant.

In a group catalog that includes galaxies of very low luminosities 
(\cite{T88} 1987, 1988) it already appeared that there may be
bound systems of dwarf galaxies.  When these group candidates were found, it
was appreciated that {\it if they are bound then they must have large
$M/L_B$ values.}  No big deal was made of these group candidates because the
statistical status of the sample was poor.  However 
if CDM hierarchical clustering theory is valid, then the deficiency of
visible dwarfs {\it requires} the existence of invisible dwarfs.  The
existence of groups of dwarfs with high $M/L$ would be a reasonable
expectation at the transition from the regime of luminous groups to the 
regime of totally invisible groups.  Hence the candidate groups of dwarf
galaxies (called `associations' in 1987) deserve renewed
attention.
With the passage of a decade there have 
been new dwarf identifications.  In fact, the
amount of new information is remarkably limited, evidence of an indirect
nature that dwarf halos with stars are not numerous.  For our purposes, the 
most 
important new surveys for dwarfs are by \cite{KK98} (1998) with follow up HI
observations by \cite{HKKE00} (2000) and the study of the Sculptor and 
Centaurus regions by \cite{CFCQ97} (1997).

Our new inventory of possible dwarf groups extends to $\sim 5$~Mpc.  Beyond
this distance extreme dwarf galaxies tend to be too faint and deficient in
HI to be reliably identified.  The search is restricted to relatively high 
Galactic latitudes since dwarfs are very difficult to find in the Galactic
plane.  In this modest volume we find four groups of 3 to 6 dwarf galaxies 
each.  One of these groups is, in fact, at the rather low latitude $b\sim18$
in a region of low obscuration near the Galactic anti-center.
The brightest galaxies in these groups have $M_B^{b,i} \sim -16$, with
$V_{circ} \sim 45$~km~s$^{-1}$.
The global properties of these small groups are summarized
in Table~1.  The numeric names of the groups are drawn from
\cite{T88} (1988).  

Before focusing on the properties of these four small groups, it is worth
a reflection on what else is going on within this 5~Mpc region.  Beyond the
Local Group there are four other groups at high galactic latitude with
big galaxies: the Canes Venatici (14-~7), M81 (14-10), Sculptor (14-13),
and Foreground Sculptor (14+13) groups (dominant galaxies: NGC~4736, NGC~3031, 
NGC~253, and NGC~55, respectively).  Information is provided in Table~1 on
these groups and also the group around M31 within the historical Local Group.
The Centaurus (14-15) group flirts
with the zone of obscuration at $b \sim 20$.
There are three more groups at $\vert b \vert < 15$:
Maffei--IC~342 (14-11), Circinus (14+20), and a newly 
revealed group around NGC~6946.  At $\vert b \vert > 30$, 80\% of galaxies 
suspected to be within 5~Mpc are in or closely associated with the groups
identified above and in Table~1.
Otherwise there are only a couple of pairs and a dozen other galaxies with 
$M_B<-14$ not associated with groups but within
the filaments called 14 and 17 (\cite{T88} 1988).  We should have a complete
census of all HI-rich systems at $\vert b \vert \gsim 20$, $M_B<-14$, and
$d<5$~Mpc.

The four dwarf groups identified in Table~1 are clearly distinguished.  
Given the small dimensions and velocity dispersions, the dwarf groups 
represent a highly significant correlation enhancement over an unclustered 
distribution.   The galaxy number densities within the r.m.s. separation shells
containing 68\% of group members ($<R_{3d}>$ in Table~1) are 3 to 80 galaxies
Mpc$^{-3}$ for the 4 groups of dwarfs.  The average number density in the
volume within 5 Mpc but excluding the groups in Table~1 is 0.1 galaxies 
Mpc$^{-3}$.

The number of high latitude
dwarf groups is comparable to the number of high latitude groups with giant
galaxies, though the number of members per group are fewer.  
The dimension, velocity dispersion, light, and inferred mass properties of the 
dwarf
groups can be compared with the properties of more familiar groups containing
large galaxies (\cite{T87} 1987).  In the summary provided in Table~1, 
projected radii $<R_p>$ are 
the mean projected separations from the geometric centers of the identified 
members with no weighting.  $<R_{3d}>$ are the equivalent 3-dimensional radii, 
directly
measured in the cases of the groups including NGC~3109 and NGC~224 (M31), but 
only derived
statistically from $(4/\pi)<R_p>$ for the cases in brackets.
Velocity dispersions $\sigma_V$ are rms differences in radial motions
from the group mean with no weighting.  Individual galaxy velocities 
for the dwarf group members are all from HI line measurements reported in 
the literature.  Since the HI profiles are 
narrow for dwarfs the
velocities are all accurate to $\pm 5$~km s$^{-1}$.

Masses are calculated
based on the `projected mass estimator' of \cite{HTB85} (1985)
\beq
M = {f_{pm} \over G(N-\alpha)} \sum_i^N R_{p,i} \Delta V_i^2 
\label{mass}
\eeq
where $f_{pm}=20/\pi$ is found by \cite{T87} (1987) to be statistically 
compatible with masses derived using the virial theorem 
(becomes $f_{pm}^{3d}=5$ and $R_p$ becomes $R_{3d}$ in the 
cases of the NGC~3109 and NGC~224 groups where three-dimensional positions are 
available),
$N$ is the number of group members, and $\alpha=1.5$ following Heisler et al.
The projected mass estimator and the mean projected radius from the group 
centroid, $R_p$, are more stable than the virial mass 
estimator and virial or harmonic radius in cases where there are close 
projections.  We make the underlying assumption that the galaxies are only
test particles in the gravitational potential well so luminosity weighting
is inappropriate and there may not be any galaxy at the actual
minimum of the potential.  The groups are expected to be bound but not
virialized so mass estimates in these non-equilibrium conditions are uncertain.
The lower mass limit that follows from the assumption the group is bound is half 
the mass given by the virial estimate.

The group including NGC~3109 is the nearest neighboring group to the Local
Group.  It is so near
that it has sometimes been considered as part of the Local Group but
galactocentric velocities are all positive and the dispersion in velocities
is tiny.  Good distances, accurate to $\sim 10\%$, are available for 5 of 6
prospective members from observations of either Cepheids or the luminosities
of stars at the tip of the red giant branch.  The remarkably similar distances
place these galaxies together and substantially beyond the Local Group
(\cite{vdb99} 1999).
The group has dimensions similar to groups with luminous galaxies
(\cite{T87} 1987) and the number density contrast of a factor of 30 over an 
average local volume of space makes it likely these galaxies are mutually bound.

Distances to the other dwarf groups are considerably less certain.  
Nevertheless,
the basic results seem well established.  Group dimensions are similar to
those of more familiar spiral groups.  Velocity dispersions are very low, 
hence inferred masses are low.  However since these are low luminosity groups,
$M/L_B$ ratios are large.  By comparison, more prominent groups have
$M/L_B=94~M_{\odot}/L_{\odot}~\pm$ factor 2 (\cite{T87} 1987; same distance
and luminosity scales).  The statistics are still slim but the groups of
dwarf galaxies seem to have $M/L_B$ values 6 times higher plus/minus a factor 2.
Mass uncertainties are large, dominated by two factors.  There are substantial
{\it random} uncertainties in velocity dispersions since sampling numbers are 
small and only line-of-sight components are observed.  Moreover, there is
ample room for {\it systematic} errors since these dwarf groups are unlikely
to be relaxed.  An envelope to random uncertainties is provided by the
factor 2 scatter in the Tully (1987) groups with 5 or more members (this 
factor 2 includes both measurement and intrinsic variances).  The random
uncertainty grows to a factor 3 if only 4 members are known.  As for systematic
error, a crude estimate of a factor 2 is provided by the difference in mass
implication between the alternative assumptions that the systems are marginally 
bound or virialized.

The group luminosities and estimated masses are plotted in Figure~6.  Groups
within 5~Mpc are indicated by the big symbols and constitute a reasonably
complete, though skimpy sample.  The triangle distinguishes the M31 group as
identified by \cite{EWGGV00} (2000).  The dwarf groups identified in this paper 
are distinguished by low estimated masses and $very~low$ luminosities.
Small symbols characterize luminous groups with $5<d<10$~Mpc, where $d$ is 
distance.

\begin{figure}[ht!]
\plotone{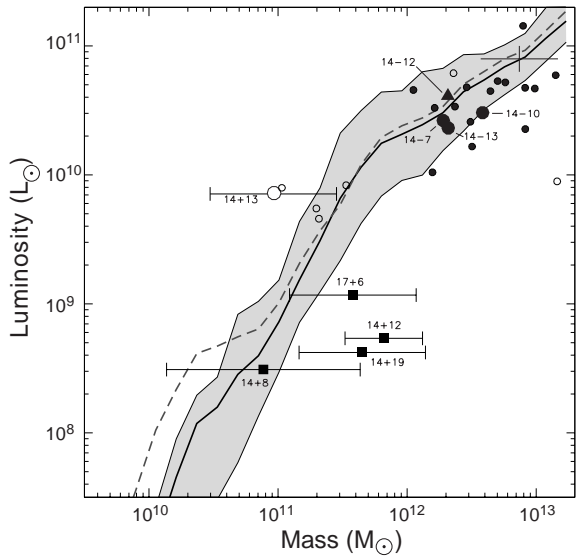}
\caption{Group mass vs. B-band group luminosity.  Filled circles: groups with 
5 or more known members identified on a basis of
luminosity density (\cite{T87} 1987).  Open circles: such groups of 3 or 4.
Filled squares: dwarf groups identified 
in this paper.  Large symbols: groups within 5~Mpc and at high galactic 
latitude.
Small symbols: other known groups within 10~Mpc.  Filled triangle: the low 
latitude but well defined M31 group.  
The mean mass and luminosity values for the sample of 49 nearby groups with 
5 or more members discused by \cite{T87} is indicated by the cross in the upper
right corner.  The horizontal arm of this cross indicates the factor 2 rms
scatter in mass at a given luminosity found in the sample of 49 groups.
The solid bold line show the mean results from semi-analytic models with the 
recipe for `squelching' as described in the text.  The grey domain includes
90\% of the model results.  The dashed line shows the mean results for
models without squelching but including supernova feedback.
}
\end{figure}

The group 14+13, including NGC 55 and NGC 300, lies in an interesting
intermediate location in Fig.~6.  This second nearest group has a very low
virial mass, in the range of the
groups of dwarfs (though uncertain by a factor of 3 since only 4 members are
identified).  
However, the 14+13, or `Foreground Sculptor' group has
two intermediate-sized galaxies so no deficiency of light.

The curves superimposed on Fig.~6 are derived from the semi-analytic modeling
with and without the photoionization squelching.  The modeling incorporates
the currently favored $\Lambda$CDM cosmology with 
$\Omega_m=0.3$, $\Omega_{\Lambda}=0.7$, $h_{75}=1$, $\sigma_8=1$, and
$\Omega_b=0.032~h_{75}^{-2}$.  The merging history of each halo is traced down
to a limiting resolution of $V_{cir} = 16$~km~s$^{-1}$, corresponding to a 
virial temperature of about $10^4$~K.  Below this temperature, gas cannot
cool via atomic processes.  In Fig.~6, curves show the mean total 
luminosities from the modeling as a function of 
halo mass, for two cases: with no squelching (dashed line) and with
squelching due to reionization at $z_{ion}=10$ (solid line).  The results of
squelched models are similar for $z_{ion} = 6-100$.  Without squelching, the
ratio of light to mass is only a weak function of mass above 
$M_{vir}\sim10^{10} M_{\odot}$, rising slightly higher
at intermediate masses.  The turndown below $M_{vir}\sim10^{10} M_{\odot}$
is mainly due to supernova feedback.  Below
$M_{vir}\sim10^{9} M_{\odot}$ there is a further cutoff because gas is not
cooled by atomic
processes.  Squelching introduces a strong cutoff beginning at masses around 
$10^{11} M_{\odot}$, about an order of magnitude above the supernova feedback
regime.

There is an important systematic that causes a displacement of the solid curve
with respect to the observed data
in Fig.~6.  The modeling pertains to {\it virialized} halos but the dwarf 
groups are {\it most unlikely to be virialized}.  The process of clustering 
in the dwarf
groups is less far along than would be the case with virialized groups of the
same mass.  One can suppose that the dwarf groups might be composed of
virialized sub-halos with masses of $1-3 \times 10^{10} M_{\odot}$ and scales
of 45-60~kpc which are in the extended process of merging.
Suppose we consider a group with mass $5\times 10^{11} M_{\odot}$.  The curve 
derived from the semi-analytic models plotted in Fig.~6 pertains to a
virialized structure and it is anticipated that $M/L$ will be modest in this
circumstance.  However a bound structure made up of several 
$3 \times 10^{10} M_{\odot}$ virialized pieces would be expected to have a
much higher $M/L$.
Suppression by reionization is greater at these very low sub-halo masses than
at the bound group mass scale.  The superposition of the semi-analytic modeling
results on the observed data is not presented as a `good fit' but rather as 
an illustration of the form a cutoff could take.  More attention needs to be 
given to the properties of bound but unvirialized structures in the 
simulations.

\section{Summary}

\noindent
1. The faint end of the luminosity function of galaxies might be 
rising in the dense environment of rich clusters but flat or falling in
the low density regions of groups.  Cold Dark Matter theory predicts that the
dark matter halo
mass function is sharply rising at the low mass end.  It seems something
is suppressing the visible manifestations of small galaxies in low density 
environments.

\noindent
2. Reionization of the universe at $z_{ion}>6$ could inhibit the collapse of 
gas in low mass potential wells for late forming galaxies.  Dynamical collapse
times inferred from the observed densities of clusters are consistent with
the picture that relatively more
dwarf halos formed {\it before} reionization in high density 
regions and relatively more formed {\it after} reionization in low density 
regions, but only if 
structure is forming at high redshift; ie, $\Omega_m \lsim 0.4$ in a flat
universe.

\noindent
3. Using semi-analytic models with a recipe for suppression of gas collapse
into low mass halos after reionization, within a $\Lambda$CDM cosmology,
it is shown that more dwarf halos formed earlier in regions that ultimately
become massive clusters.  This statement refers to dwarf halos that avoid
disruption or absorption and survive until today; many more halos formed early
and are now lost.  Qualitatively, the models anticipate that more dwarf halos 
were in place before reionization in proto-cluster environments and, compared 
with moderate density regions, the ratio of dwarf to giant galaxies should be
larger.  This fundamental expectation appears to be observed.

\noindent
4. Four small groups that only contain dwarf galaxies are found within 5~Mpc,
comparable to the number of groups that contain large galaxies.  Dynamical
evidence is found for a lot of dark matter in these groups, with 
$M/L_B \sim 300-1200~M_{\odot}/L_{\odot}$, 6 $\pm$ factor 2 times higher than in 
groups with big galaxies.  It is suggested that low mass halos which never
hosted significant star formation make up a significant fraction of the
group mass in these places.

\acknowledgments
Financial support has been provided by a NATO travel grant.

\clearpage
\onecolumn
\vspace{-2cm}

\begin{center}
\begin{tabular}{llcccccccc}
\multicolumn{10}{c}{\bf Table 1. Properties of groups within 5 Mpc}\\
\tableline
Group & Brightest & No. & Dist. & $<R_p>$ & $<R_{3d}>$ & $\sigma _V$ &
 $M$                 & $L_B$            & $M/L_B$ \\
      & galaxy    &     & Mpc   & kpc        & kpc      & km s$^{-1}$ &
 $10^{11}~M_{\odot}$ & $10^8~L_{\odot}$ & $M_{\odot}/L_{\odot}$ \\
\tableline
14$-$~7 & NGC~4736  & 22 & 4.8   & 538 & (685) & ~53 & 19.4~& 264.~ & ~~72 \cr
14$-$10 & NGC~3031  & 12 & 3.1   & 322 & (410) & 107 & 38.5~& 304.~ & ~127 \cr
14$-$13 & NGC~~253  & ~7 & 3.0   & 495 & (630) & ~69 & 20.8~& 231.~ & ~~90 \cr
14+13 & NGC~~~55  & ~4 & 1.8   & 394 & (502) & ~12 & ~0.94& ~72.~ & ~~13 \cr
\tableline
14$-$12 & NGC~~224  & 16 & 0.8   & 178 & 188 & ~77 & 20.7~& 409.~ & ~~50 \cr
\tableline
14+12 & NGC~3109  & ~6 & 1.4   & 569 &  720  & ~22 & ~6.6~& ~~5.4 & 1220 \cr
14+~8 & UGC~8760  & ~3 & 5~    & 180 & (229) & ~16 & ~0.77& ~~3.1 & ~250 \cr
14+19 & UGC~3974  & ~4 & 5~    & 356 & (453) & ~28 & ~4.5~& ~~4.2 & 1060 \cr
17+~6 & NGC~784   & ~4 & 4~    & 128 & (163) & ~36 & ~3.8~& ~11.7 & ~330 \cr
\tableline
\end{tabular}
\end{center}

\noindent
Notes to Table 1. Group memberships.

\noindent
Groups $\vert b \vert > 30$ with luminous galaxies

\noindent
Group 14$-$7: CVn~I group -- NGC~4736, NGC~4449, NGC~4244, NGC~4214, NGC~4395,
many smaller galaxies 

\noindent
Group 14$-$10: M81 group -- NGC 3031, NGC~2403, NGC~3034, NGC~3077, NGC~2366, NGC~2976, 6 others

\noindent
Group 14$-$13: Sculptor group -- NGC~253, NGC~247, NGC~7793, 4 dwarfs

\noindent
Group 14+13: Foreground Sculptor -- NGC~55, NGC~300, IC~5152, UGCA~438

-----------

\noindent
Low latitude special case

\noindent
Group 14$-$12: M31 group -- NGC 224, NGC 598, IC 10, NGC 205, NGC 221, 11 dwarfs

-----------

\noindent
Groups with only dwarfs

\noindent
Group 14+12: NGC~3109 (1.36 Mpc), Sextans~A (1.45~Mpc), Sextans~B (1.34~Mpc), 
Antlia dwarf (1.33~Mpc), GR~8~=~DDO~155 (1.51~Mpc), LSBC~D634-03 
(no distance)

For purposes of the virial analysis, D634-03 is placed at a distance
$d=<d>+<R_p>/\sqrt{2}$ where $<d>$ is the mean of the 
5 measured distances and
$<R_p>$ is the rms projected separation from the group center of the 6 
candidates.  The term $\sqrt{2}$ is the statistical correction of $<R_p>$ to the
radial direction.  

\noindent
Group 14+~8: UGC~8651, UGC~8760, UGC 8833

\noindent
Group 14+19: UGC~3755, UGC~3974, UGC~4115, KK98~65 

\noindent
Group 17+~6: NGC~784, UGC~1281, KK98~16, KK98~17

-----------

\smallskip\noindent
KK98 objects are from survey by Karachentseva \& Karachentsev (1998)

\noindent
LSBC~D634-03 is from the catalog of Schombert, Pildis, \& Eder (1997)

\end{document}